&pdflatex

\documentclass[12pt,prd,aps,amssymb,amsmath,tightenlines,showpacs]{revtex4}
\usepackage{graphicx}

\def\cP{\mathcal P}

\def\cT{\mathcal T}

\begin{document}

\topmargin=0.0cm

\title{WKB Analysis of $\cP\cT$-Symmetric Sturm-Liouville problems}

\author{Carl~M.~Bender${}^1$}
\email{cmb@wustl.edu}

\author{Hugh~F.~Jones${}^2$}
\email{h.f.jones@imperial.ac.uk}

\affiliation{${}^1$Department of Physics, Kings College London, Strand, London
WC2R 1LS, UK \footnote{Permanent address: Department of Physics, Washington
University, St. Louis, MO 63130, USA.} \\ ${}^2$Blackett Laboratory, Imperial
College, London SW7 2AZ, UK}

\begin{abstract}
Most studies of $\cP\cT$-symmetric quantum-mechanical Hamiltonians have
considered the Schr\"odinger eigenvalue problem on an infinite domain. This
paper examines the consequences of imposing the boundary conditions on a finite
domain. As is the case with regular Hermitian Sturm-Liouville problems, the
eigenvalues of the $\cP\cT$-symmetric Sturm-Liouville problem grow like $n^2$
for large $n$. However, the novelty is that a $\cP\cT$ eigenvalue problem on a
finite domain typically exhibits a sequence of critical points at which pairs of
eigenvalues cease to be real and become complex conjugates of one another. For
the potentials considered here this sequence of critical points is associated
with a turning point on the imaginary axis in the complex plane. WKB analysis is
used to calculate the asymptotic behaviors of the real eigenvalues and
the locations of the critical points. The method turns out to be surprisingly
accurate even at low energies.
\end{abstract}

\maketitle

\section{Introduction}
\label{s1}

This paper uses WKB analysis to examine the approximate solutions of complex
non-Hermitian $\cP\cT$-symmetric Sturm-Liouville eigenvalue problems on finite
domains. These problems are qualitatively different from ordinary Hermitian
eigenvalue problems because, as we will show, there is a sequence of critical
points at which the eigenvalues become pairwise complex. WKB provides an
extremely accurate asymptotic approximation to the locations of these
critical points.

A conventional Sturm-Liouville eigenvalue problem in Schr\"odinger form is a
second-order differential equation
\begin{equation}
-\psi''(x)+V(x)\psi(x)=\lambda\psi(x),
\label{e1}
\end{equation}
where $\lambda$ is the eigenvalue, accompanied by a set of homogeneous boundary
conditions
\begin{equation}
\psi(a)=0, \qquad\psi(b)=0.
\label{e2}
\end{equation}
If $a$ and $b$ are finite and the potential $V(x)$ is real and smooth for $a\leq
x\leq b$, then this eigenvalue problem is said to be a {\it regular}
Sturm-Liouville problem.

WKB theory gives a good approximation to the solution of this problem for large
eigenvalues $\lambda$ \cite{BO}. Equation (\ref{e2}) takes the form $\psi''(x)+Q
(x)\psi(x)=0$, where $Q(x)=\lambda-V(x)$, and if $\lambda\gg1$, we may assume
that $Q(x)\neq0$ on the interval $a\leq x\leq b$. Thus, there are no turning
points. When $\lambda\gg1$, the WKB approximation satisfying $\psi(a)=0$,
\begin{equation}
\psi(x)\sim\frac{C}{[Q(x)]^{1/4}}\sin\left[\int_a^x ds\,\sqrt{\lambda-V(s)}
\right],
\label{e3}
\end{equation}
is uniformly valid over the entire interval. Imposing the boundary condition
$\psi(b)=0$ then gives the eigenvalue condition
\begin{equation}
\int_a^b ds\,\sqrt{\lambda_n-V(s)}\sim n\pi\quad(n=1,2,3,\ldots).
\label{e4}
\end{equation}
For large $\lambda$ this gives an accurate asymptotic approximation to the
eigenvalues:
\begin{equation}
\lambda_n\sim\frac{n^2\pi^2}{(a-b)^2}\quad(n\to\infty).
\label{e5}
\end{equation}
Note that the eigenvalues grow like $n^2$ for large $n$. This result is general
and holds for all regular Sturm-Liouville eigenvalue problems because to
leading order in the WKB approximation the eigenvalues become insensitive to the
potential $V(x)$ and thus the eigenvalues approach those of a square-well
potential.

In this paper we study the complex $\cP\cT$-symmetric version of the eigenvalue
problem in (\ref{e1}) and (\ref{e2}). Now, instead of the potential being real,
the eigenvalue problem takes the form
\begin{equation}
-\psi''(x)-gV(ix)\psi(x)=\lambda\psi(x),
\label{e6}
\end{equation}
where $g$ is a coupling constant and $V(x)$ is a real function of its argument.
In order to respect the $\cP\cT$ symmetry, the boundary conditions are imposed
on the real-$x$ axis at parity-symmetric points:
\begin{equation}
\psi(\pm1)=0.
\label{e7}
\end{equation}
We will see that WKB provides an excellent asymptotic solution to this problem
for large $\lambda$ but that one-turning-point analysis is required.

\begin{figure}[h!]
\begin{center}
\phantom{.}\vspace{-2.5in}
\hspace{-1in}\includegraphics[scale=0.8]{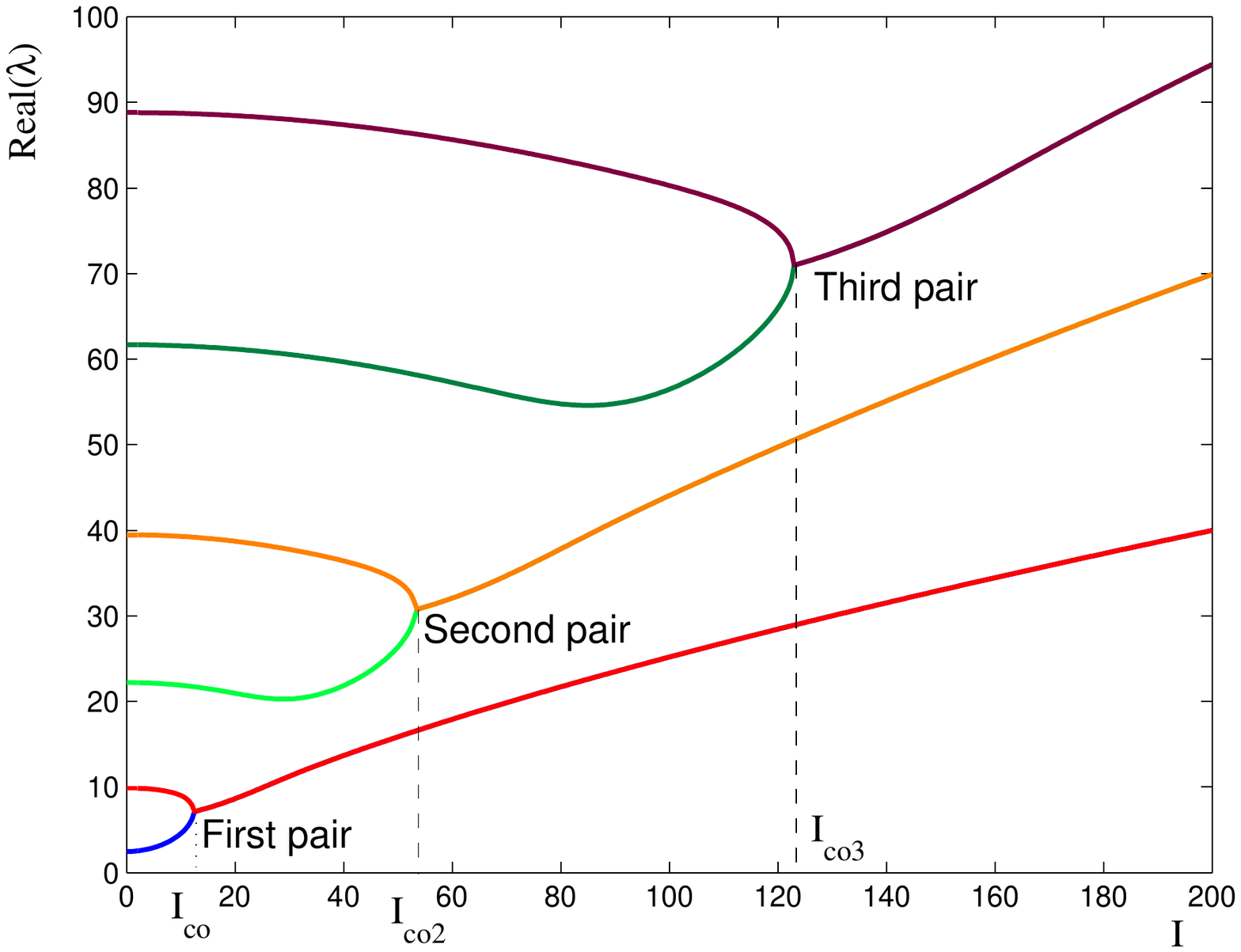}
\end{center}\vspace{-2in}
\caption{(Color on line) Real parts of eigenvalues of the $\cP\cT$-symmetric eigenvalue problem (\ref{e8}), taken from Ref.~\cite{Rub}.}
\label{F1}
\end{figure}

Eigenvalue problems like this have already been studied numerically in the
literature in a variety of physical contexts. An example of a $\cP\cT$-symmetric
Hamiltonian having a complex Sturm-Liouville eigenvalue problem like that in
(\ref{e6}) and (\ref{e7}) was first discussed by G\"unther, Znojil, and Wu
\cite{Uwe} in the context of the Squire equation of hydrodynamics. Essentially
the same Hamiltonian and boundary conditions occur in the context of
superconducting wires\cite{Rub}, and again in a different guise in the
consideration of the magnetic resonance signal of spin-polarized Rb atoms near
the surfaces of coated cells \cite{Schaden}. The Schr\"odinger eigenvalue
equation of Ref.~\cite{Rub},
\begin{equation}
-\psi''(x)-iIx\psi(x)=\lambda\psi(x),
\label{e8}
\end{equation}
is posed on the finite domain $|x|\leq 1$ with homogeneous boundary conditions
$\psi(\pm1)=0$. The eigenvalues $\lambda$ are plotted in Fig.~\ref{F1} as
functions of the real coupling constant $I$. Observe that when $0\leq I\leq
12.31$, the eigenvalues are all real. This parametric region is called the
region of {\it unbroken} $\cP\cT$ symmetry. However, as $I$ reaches the critical
value $12.31$, the two lowest eigenvalues become degenerate, and as $I$
increases past $12.31$, these eigenvalues split into a complex-conjugate pair.
Thus, for $I>12.31$ the eigenspectrum is no longer entirely real. As $I$
continues to increase, more and more pairs of real eigenvalues become degenerate
and split into complex-conjugate pairs. The critical values of a coupling
constant at which the eigenvalues become degenerate are often called {\it
exceptional points} \cite{BW}-\cite{EPe}.

\begin{figure}[h!]
\begin{center}
\includegraphics[scale=0.22, viewport=0 0 2027 644]{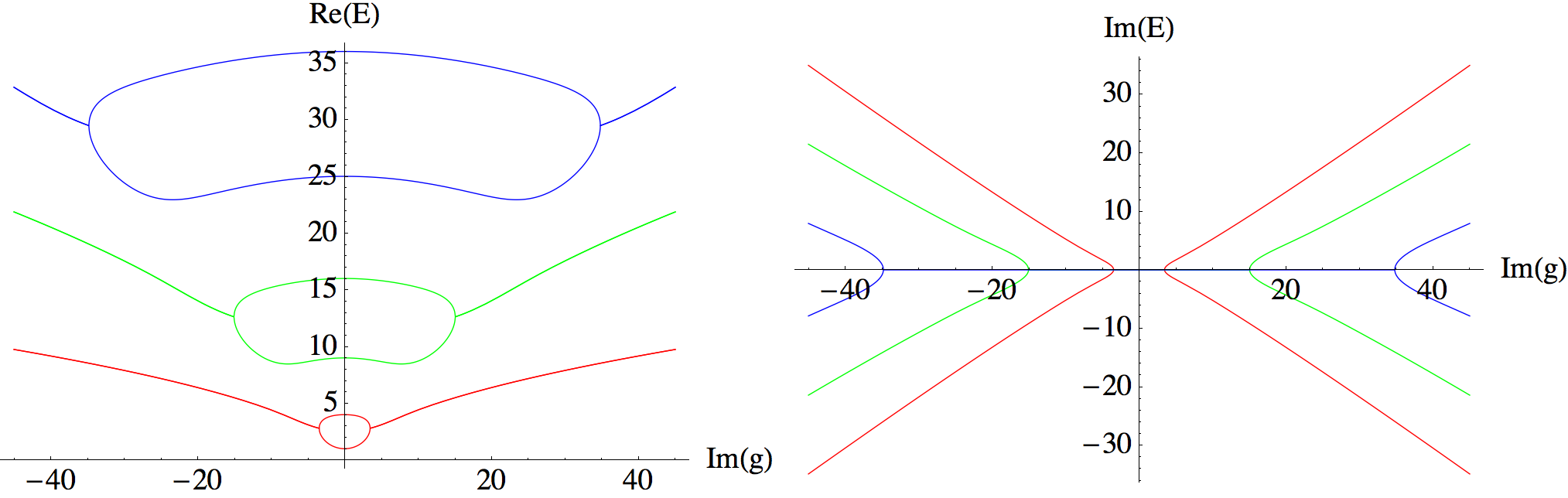}
\end{center}
\caption{Real and imaginary parts of eigenvalues of the $\cP\cT$-symmetric
Hamiltonian (\ref{e9}), taken from Ref.~\cite{BK}. The region of unbroken $\cP
\cT$ symmetry is $|g|<3.4645$.}
\label{F2}
\end{figure}

Another example of a $\cP\cT$-symmetric Hamiltonian giving rise to a
Sturm-Liouville eigenvalue problem on a finite domain is
\begin{equation}
H=-\frac{d^2}{d\theta^2}+ig\cos\theta,
\label{e9}
\end{equation}
where $g$ is a real parameter \cite{BK}. Here, the eigenfunctions are required
to be $2\pi$ periodic and odd in $\theta$. The region of unbroken $\cP\cT$
symmetry is $|g|<3.4645$ \cite{BK} (see Fig.~\ref{F2}). Note that the behavior
of the eigenvalues is virtually identical to that shown in Fig.~\ref{F1}. It is
this feature that motivated us to use WKB analysis to investigate these problems
in a more general context.

This paper is organized simply. In Sec.~\ref{s2} we describe the WKB
calculation, culminating in Eq.~(\ref{e32}), which is our principal result.
Sec.~\ref{s3} gives numerical results based on (\ref{e32}) for a variety of
Hamiltonians, including in particular those of (\ref{e8}) and (\ref{e9}).
Section \ref{s4} contains some brief concluding remarks.

\section{WKB Calculation of Eigenvalues and Critical Points}
\label{s2}

Our objective is to find accurate asymptotic approximations to the
eigenvalues $\lambda$ in (\ref{e6}) for large $\lambda$ and hence to
determine approximately the critical values of $g$. In our WKB approximation
we treat both the eigenvalues $\lambda$ and the associated values of $g$ as
large and proportional. Thus, we begin by substituting
\begin{equation}
\lambda=ag.
\label{e10}
\end{equation}
The eigenvalue equation (\ref{e6}) then takes the form
\begin{equation}
\psi''(x)=-gQ(x)\psi(x),
\label{e11}
\end{equation}
where $Q(x)=V(ix)+a$.

We will see that for large $\lambda$, the asymptotic solution to (\ref{e11}) is
controlled by a turning point on the imaginary-$x$ axis at $x=ib$, where $b$
satisfies the equation
\begin{equation}
V(-b)+a=0.
\label{e12}
\end{equation}
Near the turning point at $ib$, we write $x$ as $x=ib+yc$, where $y\ll1$. Thus,
a one-term Taylor approximation to $Q(x)$ near $ib$ is $Q(x)\sim iycV'(-b)$, and
the differential equation (\ref{e11}) is approximately
\begin{equation}
\frac{d^2}{dy^2}\psi(y)=-igc^3V'(-b)y\psi(y).
\label{e13}
\end{equation}
We then convert (\ref{e13}) to the standard Airy differential equation
\begin{equation}
\frac{d^2}{dy^2}\psi(y)=y\psi(y)
\label{e14}
\end{equation}
by setting
\begin{equation}
c=\gamma e^{i\pi/6},
\label{e15}
\end{equation}
where $\gamma=[gV'(-b)]^{-1/3}$ is real. By choosing the sign of $g$
appropriately, we may take $\gamma$ to be positive. The
relationship between the $x$ and $y$ variables is given explicitly by
\begin{equation}
y=\frac{x-ib}{\gamma}e^{-i\pi/6}.
\label{e16}
\end{equation}

Two linearly independent solutions to the Airy equation (\ref{e14}) are ${\rm
Ai}(y)$ and ${\rm Ai}(\omega y)$, where $\omega=e^{2\pi i/3}$ is a cube root of
unity. Hence, the general solution to (\ref{e14}) is
\begin{equation}
\psi(y)=K_1{\rm Ai}(y)+K_2{\rm Ai}(\omega y),
\label{e17}
\end{equation}
where $K_1$ and $K_2$ are arbitrary constants. Thus, near the turning point at
$x=ib$ on the imaginary axis, the solution $\psi(x)$ to the Schr\"odinger
equation (\ref{e11}) is
\begin{equation}
\psi(x)\sim K_1{\rm Ai}\left(\frac{x-ib}{\gamma}e^{-i\pi/6}\right)+K_2{\rm Ai}
\left(\omega\frac{x-ib}{\gamma}e^{-i\pi/6}\right).
\label{e18}
\end{equation}

Away from the turning point at $x=ib$ the solution to (\ref{e11}) can be
expressed in terms of WKB functions because $g\gg1$. When ${\rm Re}\,x<0$, the
WKB approximation to $\psi(x)$ on the left takes the form
\begin{equation}
\psi_L(x)\sim\frac{L_1}{[Q(x)]^{1/4}}\exp\left[i\int_x^{ib}ds\,\sqrt{gQ(s)}
\right]+\frac{L_2}{[Q(x)]^{1/4}}\exp\left[-i\int_x^{ib}ds\,\sqrt{gQ(s)}\right],
\label{e19}
\end{equation}
and when ${\rm Re}\,x>0$, the WKB approximation on the right is
\begin{equation}
\psi_R(x)\sim\frac{R_1}{[Q(x)]^{1/4}}\exp\left[i\int_{ib}^x ds\,\sqrt{gQ(s)}
\right]+\frac{R_2}{[Q(x)]^{1/4}}\exp\left[-i\int_{ib}^x ds\,\sqrt{gQ(s)}\right],
\label{e20}
\end{equation}
where $L_1$, $L_2$, $R_1$, and $R_2$ are arbitrary constants. Note that the
sense of integration in (\ref{e19}) and (\ref{e20}) is from left to right; in
(\ref{e19}) the integration is rightward and towards the turning point at $ib$
and in (\ref{e20}) the integration is rightward and away from the turning
point.

We must now find equations that relate the six constants in the approximations
to $\psi(x)$ in (\ref{e17}) - (\ref{e20}). Imposing the boundary condition
$\psi_L(-1)=0$ on $\psi_L(x)$ in (\ref{e19}), we obtain a condition relating
$L_1$ and $L_2$,
\begin{equation}
0=L_1\exp\left[i\int_{-1}^{ib}ds\,\sqrt{gQ(s)}\right]+L_2\exp\left[-i\int_{-1}^{
ib}ds\,\sqrt{gQ(s)}\right],
\label{e21}
\end{equation}
and imposing the boundary condition $\psi_R(1)=0$ on $\psi_R(x)$ in (\ref{e20}),
we obtain a condition relating $R_1$ and $R_2$,
\begin{equation}
0=R_1\exp\left[i\int_{ib}^1 ds\,\sqrt{gQ(s)}\right]+R_2\exp\left[-i\int_{ib}^1
ds\,\sqrt{gQ(s)}\right].
\label{e22}
\end{equation}

Four additional conditions relating the constants can be found by matching
asymptotically the WKB approximations (\ref{e19}) and (\ref{e20}) to the Airy
approximation (\ref{e17}) or (\ref{e18}) near the turning point. To perform the
match, we must show that in an overlap region near the turning point,
further asymptotic approximations to each of the asymptotic approximations that
we have already found are identical. We will need to use two
approximations to the Airy function for large argument that are valid in the
appropriate Stokes' wedges \cite{BO}:
\begin{equation}
{\rm Ai}(z)\sim\frac{1}{2\sqrt{\pi}}z^{-1/4}\exp\left(-\frac{2}{3}z^{3/2}\right)
\quad(|z|\gg1,~|{\rm arg}\,z|<\pi)
\label{e23}
\end{equation}
and
\begin{equation}
{\rm Ai}(-z)\sim\frac{1}{\sqrt{\pi}}z^{-1/4}\sin\left(\frac{2}{3}z^{3/2}+
\frac{\pi}{4}\right)\quad\left[|z|\gg1,~|{\rm arg}\,(-z)<\frac{2\pi}{3}\right].
\label{e24}
\end{equation}

We will perform the asymptotic match in the $y$ variable. Because $y=(x-ib)/c$
and $c={\rm O}\left(g^{-1/3}\right)$ is small for large $g$, it follows that
when $x$ is near $ib$, $y$ may be treated as large. Thus, it is valid to use the
asymptotic approximations (\ref{e23}) and (\ref{e24}) for the Airy functions.

\begin{figure}[h!]
\begin{center}
\includegraphics[scale=0.65]{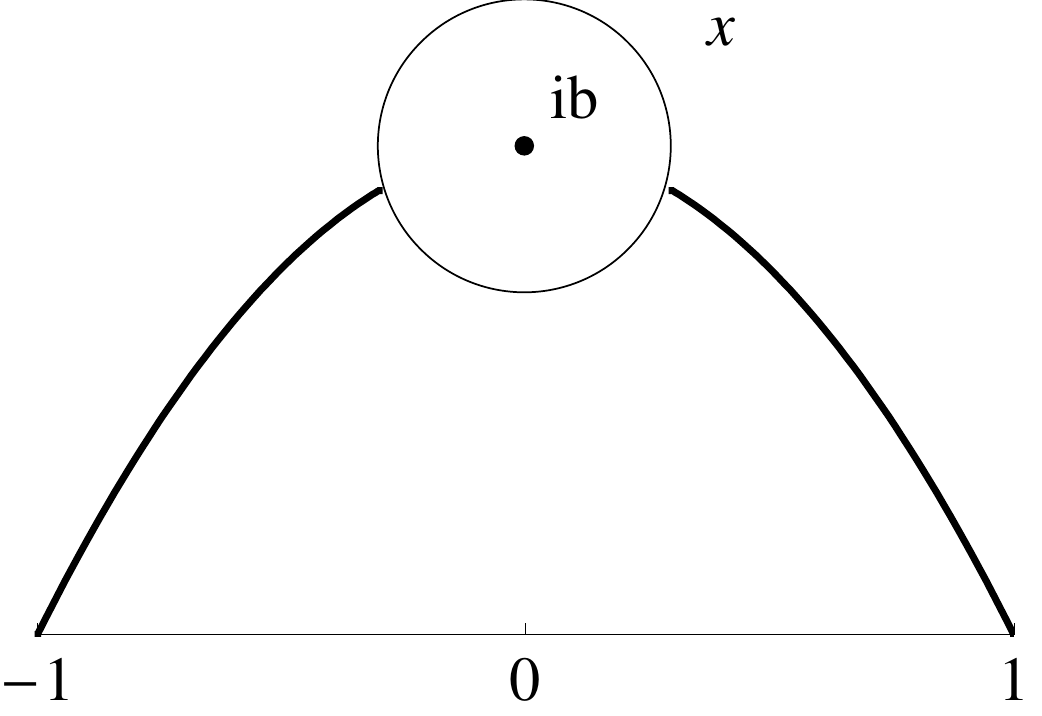}
\end{center}
\caption{Schematic path for the WKB integral from the end points $x=\pm 1$ via
the turning point at $x=ib$. Inside the circle, very close to the turning point,
the WKB approximation must be matched on either side with the asymptotic
approximation to the Airy solution.}
\label{F3}
\end{figure}

First, we examine the WKB approximation $\psi_R(x)$ in (\ref{e20}). This WKB
approximation is valid in the right-half $x$ plane away from the turning point
at $ib$; that is, outside the small circle in Fig.~\ref{F3}. From the asymptotic
approximations
\begin{eqnarray}
\int_{ib}^x ds\,\sqrt{gQ(s)} &\sim& \frac{2}{3}iy^{3/2},\nonumber\\
\left[Q(x)\right]^{-1/4} &\sim& g^{1/4}\gamma^{1/2}e^{-i\pi/6}y^{-1/4},
\label{e25}
\end{eqnarray}
the asymptotic approximation to the WKB approximation to $\psi_R(x)$ in
(\ref{e20}) becomes
\begin{equation}
\psi_R(x)\sim g^{1/4}\gamma^{1/2}e^{-i\pi/6}y^{-1/4}\left[R_1\exp\left(-
\frac{2}{3}y^{3/2}\right)+R_2\exp\left(\frac{2}{3}y^{3/2}\right)\right].
\label{e26}
\end{equation}
This approximation is valid as we approach the circle from the outside.\\

\begin{figure}[h!]
\begin{center}
\includegraphics[scale=0.55]{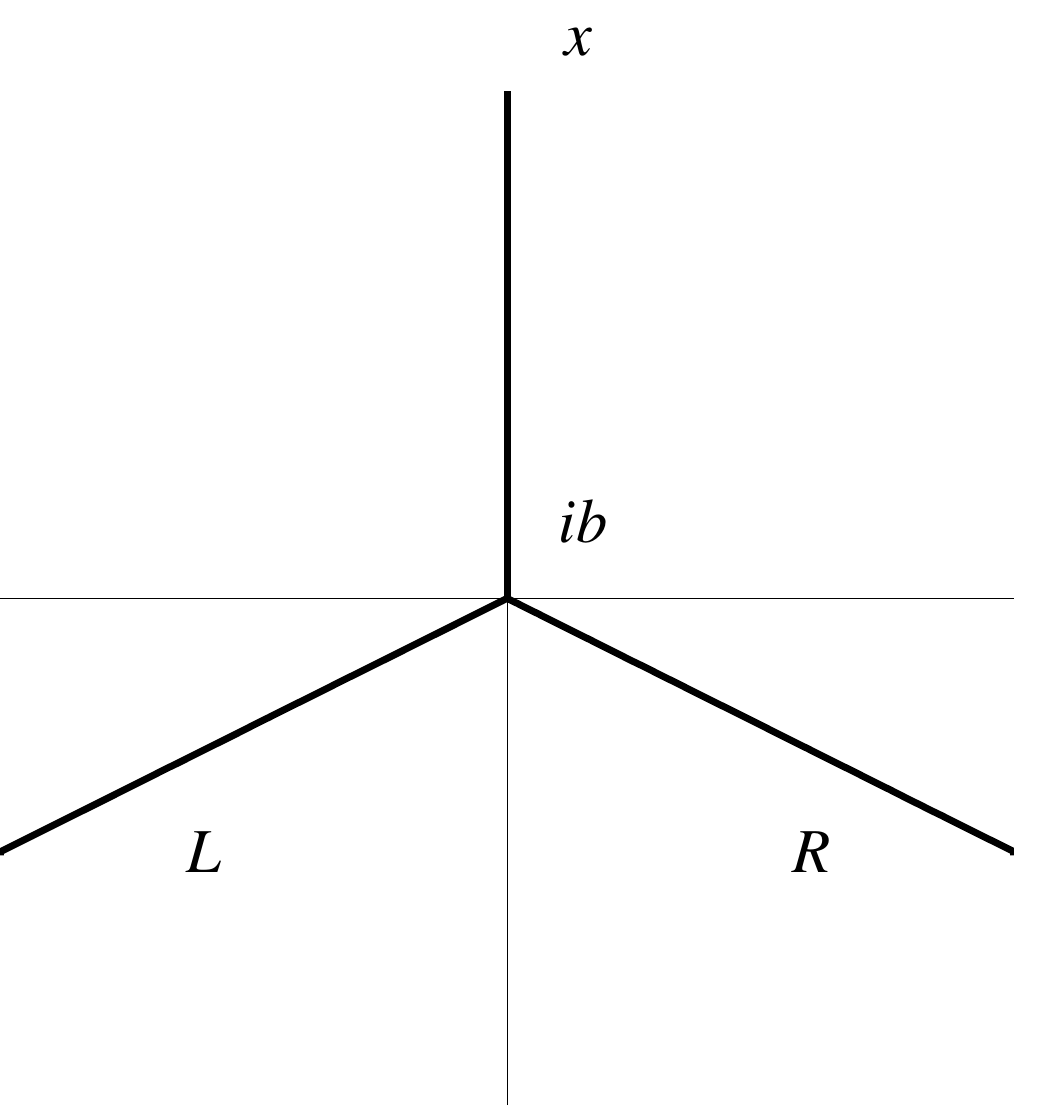}
\hspace{1.5cm}
\includegraphics[scale=0.55]{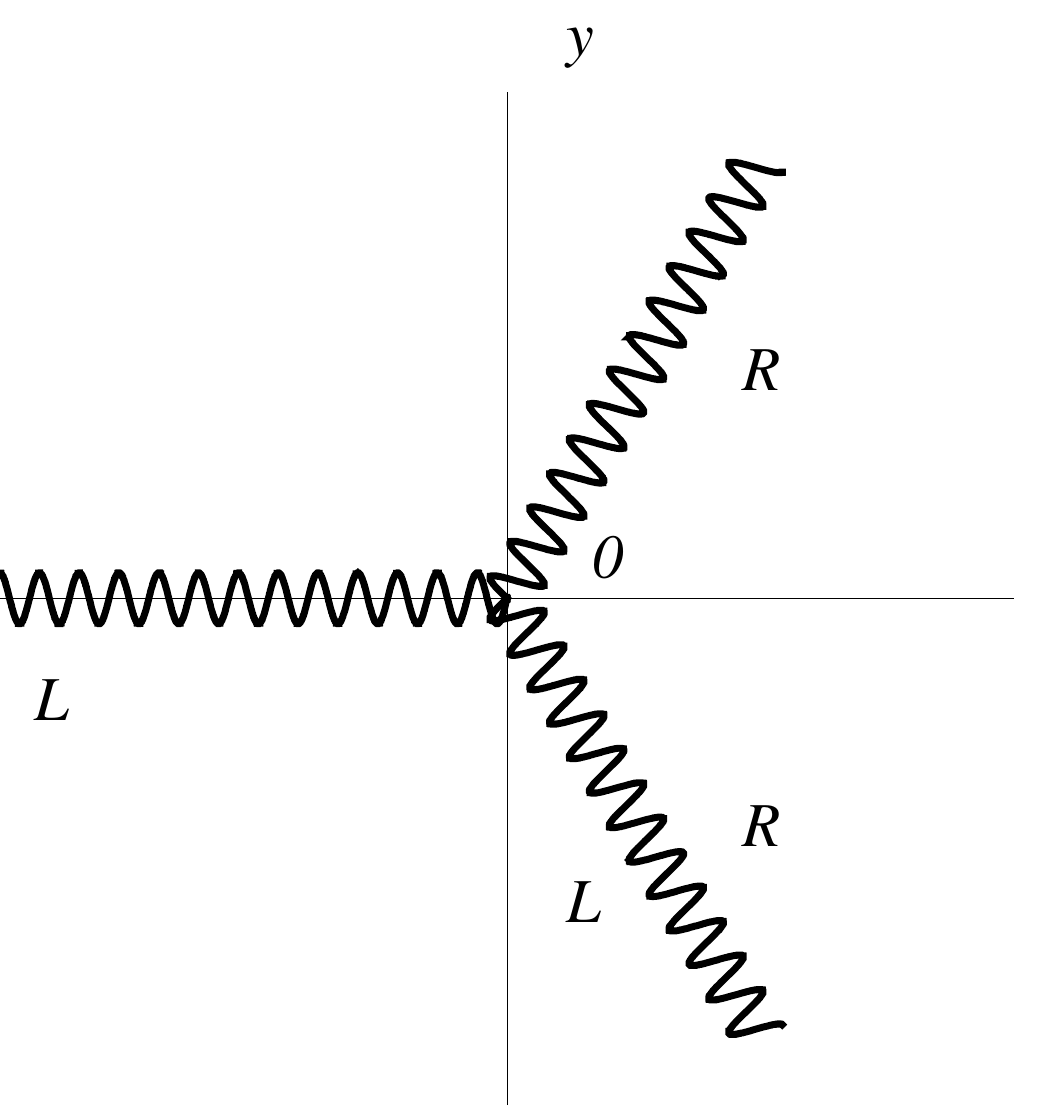}
\end{center}
\caption{Matching paths in the complex-$x$ plane (left panel) and $y$ plane
(right panel).}
\label{F4}
\end{figure}

The left panel of Fig.~\ref{F4} shows the complex-$x$ plane and the right panel
shows the corresponding complex-$y$ plane in the vicinity of the turning point
at $x=ib$. From (\ref{e16}), we see that the complex-$y$ plane is rotated by
$-30^\circ$ relative to the complex-$x$ plane. The Airy functions in (\ref{e17})
are oscillatory along the wiggly lines in the right panel.

The Airy function approximation in (\ref{e17}) is valid inside the circle in
Fig.~\ref{F3}. We must match the WKB approximation in (\ref{e26}) to the Airy
approximation along the solid line ($R$) in the left panel of Fig.~\ref{F4}.
This line corresponds to the two wiggly lines marked ($R$) in the right panel.
Because both wiggly lines satisfy the conditions of (\ref{e23}), as we approach
the right edge of the circle from inside, we use only this asymptotic
approximation to obtain
\begin{equation}
\psi(x)\sim\frac{1}{2\sqrt{\pi}}y^{-1/4}\left[K_1\exp\left(-\frac{2}{3}y^{3/2}
\right)+K_2\exp\left(\frac{2}{3}y^{3/2}-\frac{i\pi}{6}\right)\right].
\label{e27}
\end{equation}
This asymptotic match produces two algebraic equations for the coefficients:
\begin{eqnarray}
\frac{K_1}{2\sqrt{\pi}}&=&R_1 g^{1/4}\gamma^{1/2}e^{-i\pi/6},\nonumber\\
\frac{K_2}{2\sqrt{\pi}}&=&R_2 g^{1/4}\gamma^{1/2}.
\label{e28}
\end{eqnarray}

Next, we further approximate the WKB approximation in (\ref{e19}). Again, using
the formulas in (\ref{e25}), we find that in the left-half $x$ plane
\begin{equation}
\psi_L(x)\sim g^{1/4}\gamma^{1/2}e^{i\pi/3}(-y)^{-1/4}\left[L_1\exp\left(\frac
{2}{3}i(-y)^{3/2}\right)+L_2\exp\left(-\frac{2}{3}i(-y)^{3/2}\right)\right].
\label{e29}
\end{equation}
For the Airy approximation  (\ref{e17}) we must now perform the asymptotic
match along the solid line ($L$) in the left panel of Fig.~\ref{F4}. This line
corresponds to the two wiggly lines marked ($L$) in the right panel. In this
case the wiggly line along the negative-$y$ axis requires that we use the
asymptotic approximation (\ref{e24}) for the Airy function ${\rm Ai}(y)$. For
the other Airy function ${\rm Ai}(\omega y)$ we use the asymptotic approximation
(\ref{e23}). This allows us to obtain the following asymptotic approximation to
the Airy-function approximation to $\psi(x)$ in (\ref{e17}):
\begin{equation}
\psi(x)\sim\frac{K_1}{\sqrt{\pi}}(-y)^{-1/4}\sin\left[\frac{2}{3}(-y)^{3/2}+
\frac{\pi}{4}\right]+\frac{K_2}{2\sqrt{\pi}}(-y)^{-1/4}e^{i\pi/12}\exp\left[
\frac{2}{3}i(-y)^{3/2}\right].
\label{e30}
\end{equation}
This asymptotic match produces two further algebraic equations
\begin{eqnarray}
\frac{K_1}{2\sqrt{\pi}}&=&L_2g^{1/4}\gamma^{1/2}e^{-i\pi/6},\nonumber\\
\frac{1}{2\sqrt{\pi}}\left(-K_1e^{2i\pi/3}+K_2\right)&=& L_1g^{1/4}\gamma^{1/2}.
\label{e31}
\end{eqnarray}

Finally, we combine all six algebraic equations (\ref{e21}), (\ref{e22}),
(\ref{e28}), and (\ref{e31}), and obtain a secular equation that determines the
eigenvalues:
\begin{equation}
\sin\left[\int_{-1}^1 ds\,\sqrt{gQ(s)}\right]+\frac{1}{2}\exp\left[i\int_{-1}^{i
b}ds\,\sqrt{gQ(s)}-i\int_{ib}^1 ds\,\sqrt{gQ(s)}\right]=0.
\label{e32}
\end{equation}
Equation (\ref{e32}) is our main result. This equation is {\it real}, which is a
general feature of all $\cP\cT$-symmetric secular equations \cite{BBBM}. To see
that it is indeed real note that the argument of the sine can be written as $2
{\rm Re}I_+$, while that of the exponential can be written as $2\ {\rm Im} I_+$,
where $I_+=\int_{ib}^1 ds\ \sqrt{gQ(s)}$

The first term of (\ref{e32}) is what one would obtain from a path going
directly along the real axis from $x=-1$ to $x=1$ without going through the
turning point at $x=ib$. This corresponds to the no-turning-point result of
(\ref{e3}) for the Hermitian case. We shall see in the next section that this
term by itself gives very little structure. The second term is the exponential
of a generically large real number. When that number is large and negative it
makes very little difference to the calculation of the eigenvalue, and when it
is large and positive the equation has no real solutions because $|\sin\theta|<
1$. All the interesting structure, including the critical points, comes from the
interplay of the two terms in the region where the exponent passes through a
zero.
\section{Numerical calculations}
\label{s3}
\begin{figure}[h!]
\begin{center}
\includegraphics[scale=0.74]{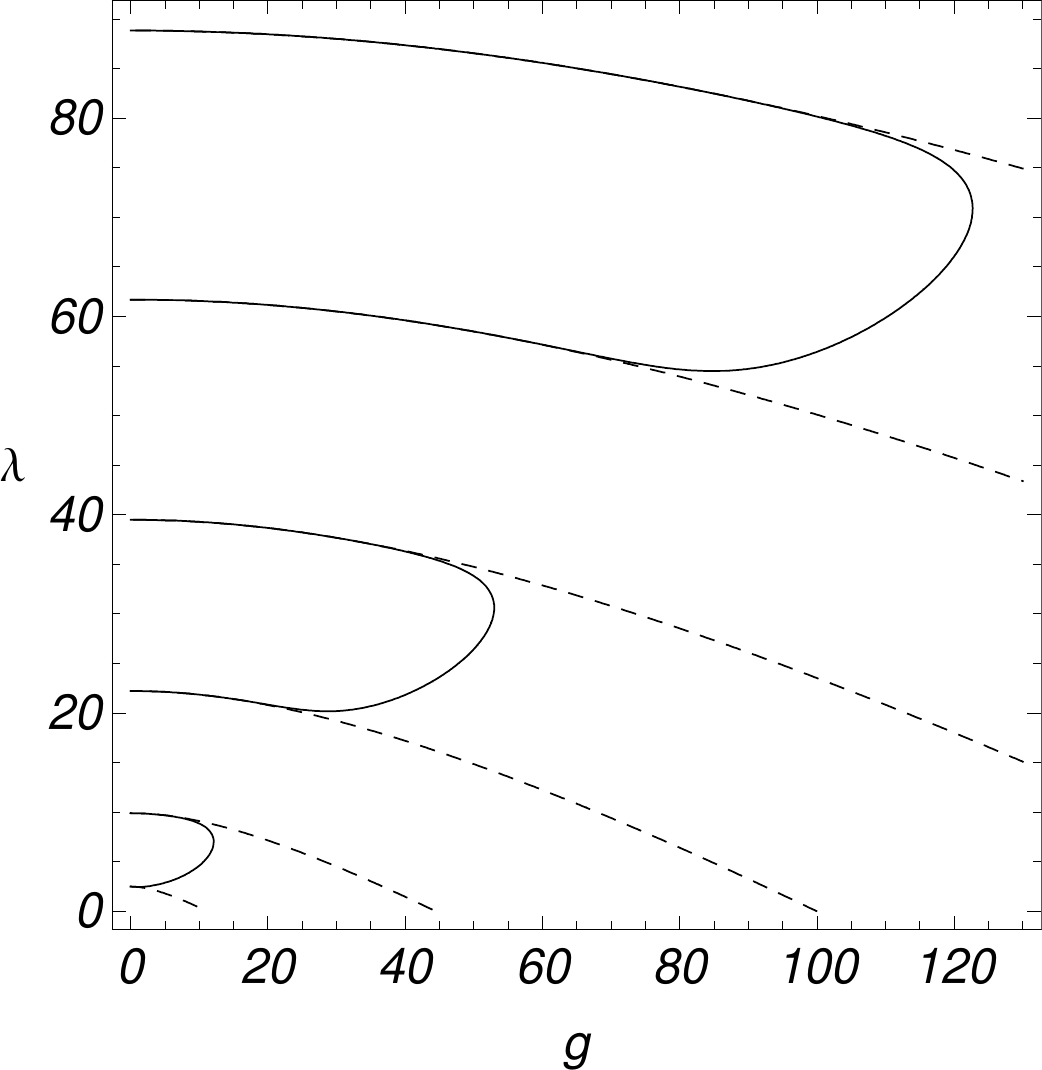}
\end{center}
\caption{WKB approximation for the real energy levels of the Airy potential $V(i
x)=ix$. The no-turning-point approximation (dashed line), which is obtained by
neglecting the second term of (\ref{e32}), exhibits no critical points. The full
result (solid line), which includes the second term, indeed gives the critical
points as a consequence of the interplay of the two terms. The solid line is
essentially indistinguishable from the numerical results in Fig.~\ref{F1}, even
at low energies.}
\label{F5}
\end{figure}

In Fig.~\ref{F5} we show the predictions of (\ref{e32}) for the real eigenvalues
of the Airy potential $V(ix)=ix$ occurring in (\ref{e8}) (with $g$ taking the
role of $I$). These are the solid lines, which are essentially indistinguishable
from the numerical results of Fig.~\ref{F1}. Remarkably, the range of validity
of the WKB approximation extends to small values of $g$ and $\lambda$. In
Fig.~\ref{F5} we also show as dotted lines the result of the no-turning-point
WKB approximation, namely the first term of (\ref{e32}). While this
approximation reproduces correctly the square-well eigenvalues for $g=0$, it
fails to reproduce the interesting structure, which arises as a result of the
interplay between both terms of (\ref{e32}). For this potential the full WKB
approximation predicts that the critical points occur at $\lambda=g/\sqrt{3}$,
which is actually an exact result \cite{Uwe,Shk}.

\begin{figure}[h!]
\begin{center}
\includegraphics[scale=0.74]{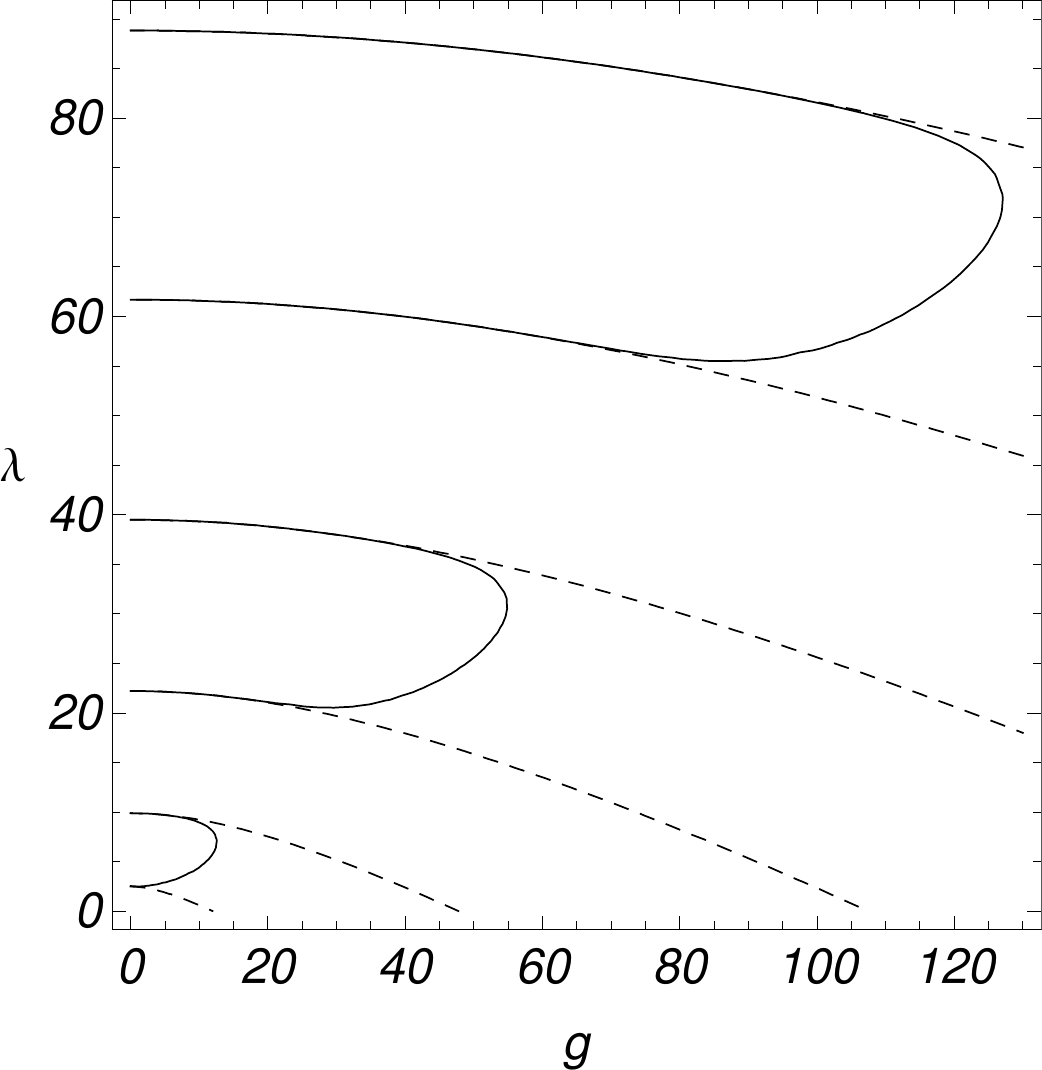}
\end{center}
\caption{WKB approximation for the real energy levels of the potential $V(ix)=i
\sin{x}$. The notation is as in Fig.~\ref{F5}. The solid line is essentially
indistinguishable from the numerical results in Fig.~\ref{F2}, even at low
energies.}
\label{F6}
\end{figure}

In Fig.~\ref{F6} we show the analogous results for the sinusoidal potential in
(\ref{e9}) (with $\theta=\pi/2-x$). The same features of the approximation are
true here too, and (\ref{e32}) reproduces very closely the numerical results of
Fig.~\ref{F2}. The reason that the spectra of the two potentials in (\ref{e8})
and (\ref{e9}) are so close is that within the range of the integrals $\int ds
\sqrt{gQ(s)}$ occurring in (\ref{e32}) the function $\sin{s}$ is well
approximated by $s$. Note that in this case $b=\arcsin{a}$ is multivalued, so
there is a series of turning points on the imaginary axis. For the WKB
calculation we considered only the nearest turning point, which is close to
$ia$.

\begin{figure}[h!]
\begin{center}
\includegraphics[scale=0.74]{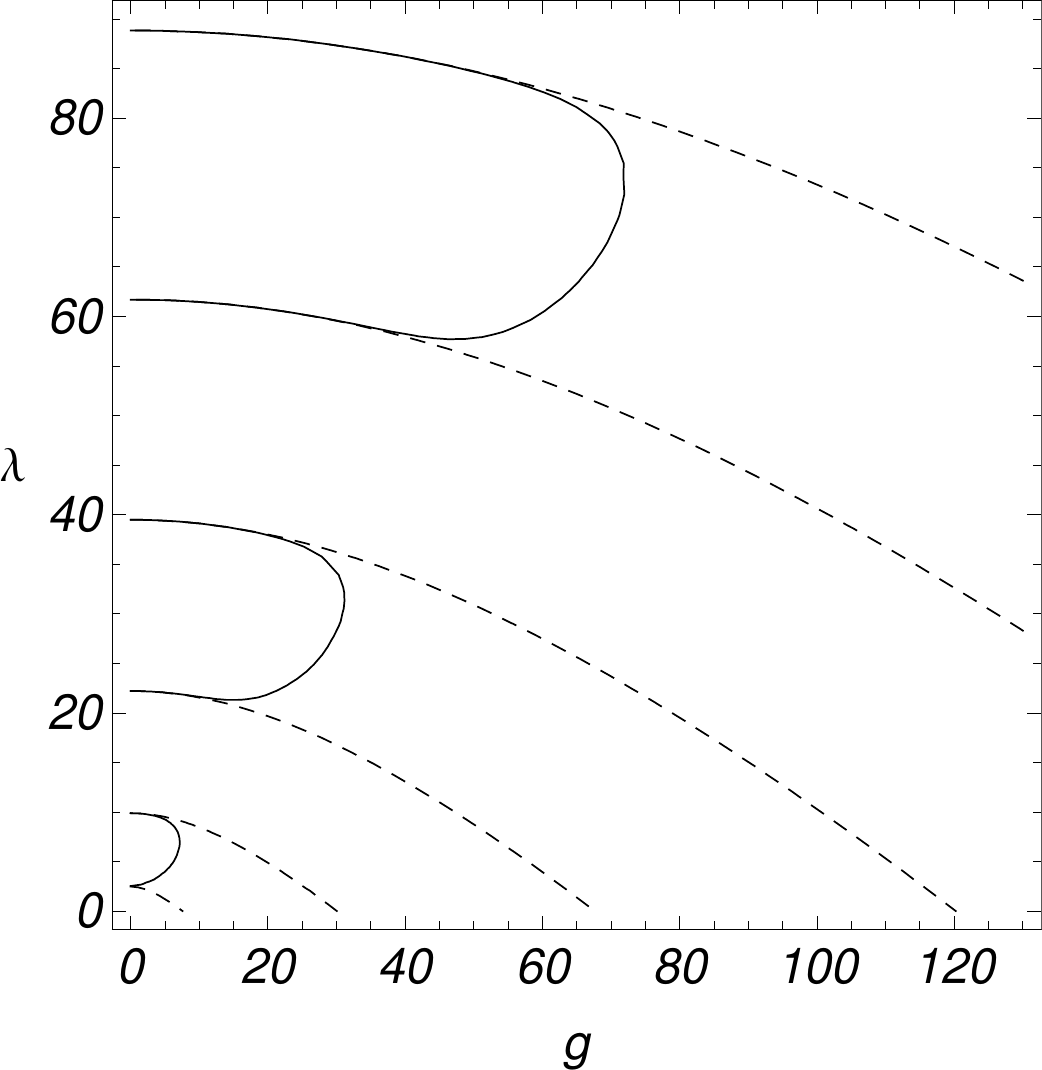}
\end{center}
\caption{WKB approximation for the real energy levels of the potential $V(ix)=i
\sin(2x)$. These are distinct from those of the scaled Airy potential $V(ix)=2ix
$.}
\label{F7}
\end{figure}

In Fig.~\ref{F7} we show the analogous results for the potential $V(ix)=i\sin(2x
)$. Again, these are indistinguishable from numerical results obtained from a
shooting method, but now they differ from the eigenvalues of the scaled Airy
potential $V(ix)=2ix$, which demonstrates that within the range of the integrals
$\int ds \sqrt{gQ(s)}$ it is not valid to neglect the $s^3$ term in the
expansion of $\sin(2s)$.

\begin{figure}[h!]
\begin{center}
\includegraphics[scale=1.2]{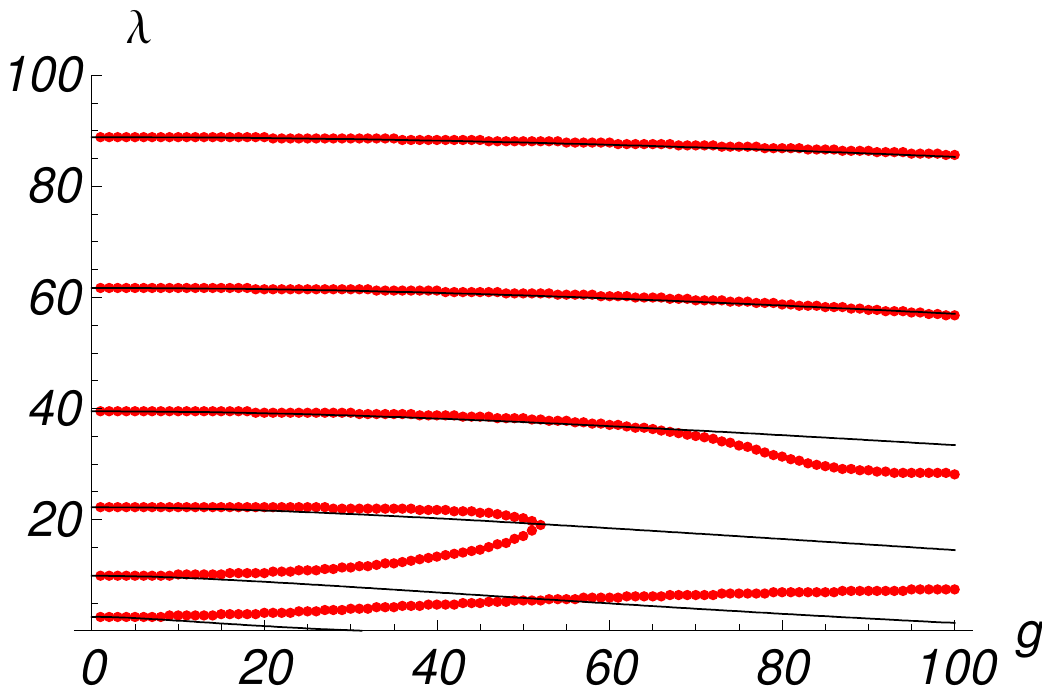}
\end{center}
\caption{(Color on line) WKB approximation for the real energy levels of the
potential $V(ix)=-ix^3$ compared with the numerical results (red dots). The
numerical results show just one low-energy critical point in the range of $g$
considered, while the WKB approximation gives no critical points. This is
because the second term of (\ref{e32}) is negligible in this case.}
\label{F8}
\end{figure}

What happens if we apply our WKB approximation to the potential $V(ix)=-ix^3$?
Now there are three complex turning points, which are located at $x=ib\{1,\omega
,\omega^2\}$. Since our WKB analysis involved only one turning point, we chose a
path going through the turning point at $ib$. The numerical results in
Fig.~\ref{F8} show a single low-energy critical point, which the WKB
approximation fails to reproduce. In this case the second term of (\ref{e32}) is
always negligible, so there is no effective interplay between the two terms. On
the other hand, the first term tracks very well the curves for the higher energy
levels. It is conceivable that a WKB path going through the other two turning
points would reproduce the low-energy structure, and we intend to address this
problem in a future publication.

\section{Comments and discussion}
\label{s4}
In this paper we have used WKB analysis to derive the simple formula
(\ref{e32}), which gives an extremely accurate approximation for the energy
levels $\lambda$ of $\cP\cT$-symmetric Sturm-Liouville eigenvalue problems. This
derivation requires the use of one-turning-point analysis. If a no-turning-point
analysis is used one still obtains a good approximation to the high-lying energy
levels for fixed coupling constant $g$. However, the no-turning-point analysis
is unable to reproduce the critical points that occur when $g$ and $\lambda$ are
both large. What is most remarkable is that the one-turning-point formula gives
an accurate description of the spectrum and critical points even when $g$ and
$\lambda$ are not large.

Particular examples that we have studied included the $ix$ and $i\sin{x}$
potentials. In these cases we were able to understand the close equality of
their respective spectra and why the spectra of the pair of potentials $2ix$ and
$i\sin(2x)$ were not equal. In the case of the $ix^3$ potential the WKB
approximation correctly reproduces that higher energy levels, which do not
exhibit any critical behavior. There is, however, a low-energy critical point,
as seen in Fig.~\ref{F8}, which our WKB approximation does not reproduce. It is
an open question whether a two-turning-point approximation would be able to do
so.

\begin{acknowledgments}
We thank U.~G\"unther and Z.~H.~Musslimani for useful discussions. CMB is
supported by the U.K.~Leverhulme Foundation and by the U.S.~Department of
Energy.
\end{acknowledgments}

\end{document}